# Vortex pinning in high-$T_c$ materials via randomly oriented columnar defects, created by GeV proton-induced fission fragments

J. R. Thompson,[a,b,*] J. G. Ossandon,[c] L. Krusin-Elbaum,[d] H. J. Kim,[b] K. J. Song,[b] D. K. Christen,[a] and J. L. Ullmann[e]

[a] Oak Ridge National Laboratory, Oak Ridge, Tennessee 37831-6061 USA
[b] Department of Physics, University of Tennessee, Knoxville, Tennessee 37996-1200 USA
[c] Department of Engineering Sciences, University of Talca, Curico, Chile
[d] IBM Watson Research Center, Yorktown, New York 10598-0218
[e] Los Alamos National Laboratory, Los Alamos, New Mexico 87545

**Abstract**

Extensive work has shown that irradiation with 0.8 GeV protons can produce randomly oriented columnar defects (CD's) in a large number of HTS materials, specifically those cuprates containing Hg, Tl, Pb, Bi, and similar heavy elements. Absorbing the incident proton causes the nucleus of these species to fission, and the recoiling fission fragments create amorphous tracks, i.e., CD's. The superconductive transition temperature $T_c$ decreases linearly with proton fluence and we analyze how the rate depends on the family of superconductors. In a study of Tl-2212 materials, adding defects decreases the equilibrium magnetization $M_{eq}(H)$ significantly in magnitude and changes its field dependence; this result is modeled in terms of vortex pinning. Analysis of the irreversible magnetization and its time dependence shows marked increases in the persistent current density and effective pinning energy, and leads to an estimate for the elementary attempt time for vortex hopping, $\tau \sim 4 \times 10^{-9}$ s.

**PACS codes:** 74.60.Ge, 74.62.Dh, 74.25.Ha, 74.72.Fq

**Keywords:** random columnar defects, vortex pinning, magnetization; thallium cuprates ($Tl_2Ba_2CaCu_2O_x$ )

[*] Corresponding author:
J. R. Thompson
Postal address: Oak Ridge National Laboratory, PO Box 2008, Oak Ridge, Tennessee 37831-6061 USA
Phone:      001 (865) 574-0412
Fax:         001 (865) 574-6263
Email:       JRT@UTK.EDU

**Introduction**

For many applications envisioned for high temperature superconductors, the materials must conduct large electric currents in high magnetic fields. To obtain the necessary pinning of vortices, some form of correlated disorder [1] has proven to be most effective. In the first generation of HTS advanced conductors, Bi-cuprates were textured and fabricated with adequate texture to allow good intergrain current transport. Unfortunately, the Bi-based compounds (like many HTS materials) have particularly rapid, thermally activated movement of vortices and a low lying irreversibility line $H_{irr}(T)$ where the persistent current density $J$ becomes too small to measure. To suppress thermal activated motion of vortices, strong vortex "pins" have been fabricated; especially useful are columnar defects, which can be created by heavy ion irradiation [2] or by fission fragments generated internally by irradiation with GeV protons[3]; the latter process produces randomly oriented, "splayed" tracks. Somewhat similar defect configurations can be formed by neutron irradiation of HTS's containing adequately dispersed uranium nuclei, which are caused to fission by absorption of a neutron. [4] Both parallel and splayed configurations of columnar defects are effective in enhancing $J$, sometimes by orders of magnitude, and in elevating the irreversibility line. However, heavy ions generally penetrate materials only to the depth of few tens of micrometers, constraining potential technological implementation.

To overcome the limitations imposed by the small penetration of heavy ions, we have utilized fission-inducing GeV protons. Protons of 0.8 GeV protons are very penetrating with a range of ~ 0.6 meters in a high-$T_c$ material. In this process, an incident proton is absorbed by some heavy nuclei, e.g., Bi, Hg, Tl, Pb, ..., that are components of many high-$T_c$ superconductors. In these high-Z nuclei, the cross-section $\sigma_f$ ~100 millibarns for prompt fission is moderately large; $\sigma_f$ increases approximately exponentially with the quantity $Z^2/A$, where $Z$ is the atomic number and $A$ is the atomic mass. Upon entering a heavy nucleus, the energetic proton causes spallation of several neutrons and protons, which is the process used in spallation neutron sources. The highly excited nucleus immediately fissions into two fragments having roughly half of the mass and charge of the parent nucleus, with an energy of ~100 MeV per fragment. These energetic recoiling nuclear fragments create columnar defects – latent track - deep inside bulk material. The number density of fission events is given by

$$(N/V)_{fission} = \sigma_f\ (N/V)_{nuclei}\ \varphi_p \quad (1)$$

Here $(N/V)_{nuclei}$ is the density of fissionable nuclei, e.g., Bi, which is obtained from the material's stoichiometry and lattice parameters, and $\varphi_p$ is the number of incident protons per unit area. One can convert $(N/V)_{fission}$ into an approximate areal density of defects by multiplying it by the length $\ell \approx 8$ μm of the fission track; finally, this is conveniently reexpressed as an equivalent vortex density - a 'matching field' $B_\Phi$ - by multiplying by the flux quantum $\Phi_0$.

The fission process is nearly random in direction, and consequently the tracks are randomly distributed in angle. This random splay and a distribution of track diameters in the range ~ 4-8 nm have been imaged by transmission electron microscopy in Bi-cuprate and Hg-cuprate superconductors. These images provide visual evidence that the fragment's energy is sufficiently high to locally amorphize the material.

The defects help to localize vortices by a local suppression of the order parameter in the superconductor, thereby creating a potential well for a vortex line or "pancake" vortex. At the same time, the overall order parameter suffers some depression, which reduces the superconductive transition temperature $T_c$. In Fig. 1 is shown the dependence of $T_c$ on the fluence of 0.8 GeV protons, for several different families of superconductors, including two separate preparations of Tl-2212 with somewhat different initial values for $T_c$. In all cases, the $T_c$ decreases linearly with fluence. As suggested by Eq. 1, we find experimentally that the decrease, $\Delta T_c$, depends on the density of defects produced, rather than the proton fluence itself. This is illustrated in the inset of Fig. 1, where the depression rate $-\Delta T_c/\Delta\varphi_p$ is proportional within experimental uncertainties to the product of $\sigma_f (N/V)_{nuclei}$ that appears in Eq. 1. A linear depression of $T_c$ due to irradiation with medium mass and heavy ions has been analyzed by Zhu et al. [5] using a thermal spike model. They find for $YBa_2Cu_3O_{7-\delta}$ materials that the change in $T_c$ depends not only on the particle fluence, but also on other factors. These include the electronic stopping power ($-dE/dx$) for energetic ions; for the materials in Fig. 1, however, the stopping power varies little ($\pm 7\%$) and does not significantly affect the proportionality shown in the inset of Fig. 1. Zhu et al. also argue that the efficiency for forming columnar defects depends on the thermal diffusivity of the various compounds at high temperatures. Differences in this quantity, which is not available experimentally, may account for the residual scatter in the data in the inset.

**Experimental results**

The first materials to be modified by splayed columnar defects, created by GeV proton-induced fission fragments, were Bi-cuprates [3]. This work demonstrated that the persistent current density, as determined magnetically, could be increased greatly. In addition, it showed that the irreversibility line could be elevated in temperature by 20 K or more, thereby increasing significantly the temperature range for potential applications. Complementary studies of the transport current density provided conclusive proof that the added defects increased both the intra- and inter-grain current density in prototype Bi-cuprate/Ag tapes [6]. In subsequent studies, it was shown that enhanced levels of current density are accompanied by and intimately related with a strong suppression of thermally activated vortex motion. In these flux creep studies, the temporal stability of $J$ was investigated by measurement of $J$ versus time at fixed field and temperature. Interestingly, the temperature dependent decay of current decreases markedly, but there remains a temperature *independent* decay with non-trivial rates $S = -d\ln(J)/d\ln(t)$ lying near 0.03 or greater [7]. The decay has been attributed to quantum tunneling of vortices between pinning sites, rather than thermal activation over pinning barriers, and the rate is large enough to reduce the realized current density to levels well below those expected for the critical current density.

Continuing studies demonstrated that substantial enhancements in low-loss current density could be achieved by fission-generated defects in many families of HTS materials, in various morphologies [8]. These include films and bulk polycrystalline Hg-cuprates [9] and Tl-based cuprate materials. To explore the impact of mass anisotropy on the properties of HTS's containing randomly oriented columnar defects, we examined the current density and normalized current decay rate $S$ in the series Hg-1021, Hg-1212, and Hg-1223, with mass anisotropies $\gamma$ near 25, 50, and 60, respectively. As the anisotropy increased, the superconductor recovered features [10,11] very similar to those observed in YBCO with *parallel* columnar defects [12]: in particular, the creep rate $S(T)$ undergoes a large peak near $T_c/2$ that arises from variable range hopping (VRH) of vortices, in magnetic fields with vortex density $B << B_\Phi$. Physically, the recovery of VRH arises from an effective focusing of the field and the defect array into an angular range near the c-axis. Thus the Hg-1223 material with the highest mass anisotropy resembled most closely the case with parallel defects in YBCO.

In other work, we have compared and contrasted the vortex pinning produced in Hg-cuprates by GeV proton-induced defects with that provided by fast neutron irradiation and chemical doping [13]. In general, the correlated disorder provided by splayed columnar defects led to a higher lying irreversibility line and other desirable features

Recent studies of $Tl_2Ba_2CaCu_2O_x$ materials illustrate some of the significant changes that can occur in the equilibrium and irreversible superconductive properties, upon doping with fission-generated columnar defects [14]. Proton fluences of 0 to $8.7\times10^{16}$ ions/cm$^2$ were used, corresponding to nominal matching fields from 0 to ~1.5 T. The associated damage reduced the superconductive transition temperature $T_c$ from 97 to 87 K, as seen in Fig. 1, and reduced the Meissner fraction (field-cooled magnetization) from 58 % to 49 % of ideal diamagnetism. Let us now consider how vortex pinning by random CD's affects the equilibrium magnetization. In the mixed state of an ideal, pin-free superconductor, vortex-vortex interactions contribute to the energy of the system in a way that depends on the vortex density $B$. This leads in London theory to the well known logarithmic field dependence for the equilibrium magnetization in the limit that $H_{c1} \ll H \ll H_{c2}$, with [15]

$$M_{eq} = -(\varepsilon_o/2\Phi_o)\times\ln(\eta H_{c2}/B). \qquad (2)$$

Here $\varepsilon_o = (\Phi_o/4\pi\lambda_{ab})^2$ is the line energy and $\lambda_{ab}$ is the London penetration depth for supercurrents flowing in the *ab*-plane. Figure 2 shows $M_{eq}$ versus applied field $H \cong B$ on a logarithmic scale, for several materials at $T = 40$ K; the background-corrected values were obtained from hysteresis loops by averaging the moments measured in increasing- and decreasing-field history. From the Bean critical state model, this gives a very close approximation to $M_{eq}$ when the hysteresis is small (or zero). For the virgin Tl-2212 with = 0, the linear dependence in Fig. 2 shows that the simple London relation describes the system quite well. From the slopes of data like these, the temperature dependent penetration depth can be obtained [14].

With the addition of randomly oriented columnar defects, however, $M_{eq}$ decreases significantly in absolute magnitude and becomes multiply valued, as is evident in Fig. 2. With a small defect density, e.g., $B_\Phi = 0.18$ T, $M_{eq}$ is affected more strongly at low fields, where most vortices can reside on CD's and the system energy changes significantly. Then at higher fields where all CD's are filled, the addition of more vortices resembles the case with the unirradiated superconductor and the values for $M_{eq}$ become similar. Still higher defect densities, e.g., $B_\Phi = 1 - 1.5$ T, lie in the middle of the experimentally accessible field range and they lead to the profound reduction in the magnitude of the equilibrium magnetization observed in Fig. 2.

To better understand the underlying physics, we use the vortex-defect interaction model of Wahl et al. [16,17]. Their analysis provides that

$$M_{eq} = -(\varepsilon_o/2\Phi_o)\times\ln(\eta H_{c2}/B) - (U_o/\Phi_o)\{1-[1+U_oB_\Phi/\varepsilon_oB]\exp(-U_oB_\Phi/\varepsilon_oB)\} \qquad (3)$$

The first term is the usual London contribution and the second term represents the modification produced by CD's with pinning energy $U_o$ and average density $B_\Phi$. The correction term, which is significant for intermediate fields, vanishes in large fields $B \rangle\rangle B_\Phi$. This theory was developed for the case of the magnetic field applied along parallel tracks. By comparison, the present Tl-2212 materials are considerably more complex in real space, with distributions of both crystallographic orientations and CD directions. However, as discussed above for the analogous Hg-cuprate superconductors, high superconductive anisotropy leads to an effective 'refocusing' toward the c-axis for both the defect and vortex arrays. Mindful of this complexity, we ask whether a reasonable quantitative description can be obtained using the theory. Thus we make no assumptions about temperature dependencies or interrelationship of the variables, but rather we vary manually the parameters in Eq. 3 to describe $M_{eq}(H)$ at each temperature. The resulting modeling is reasonably successful, as shown by the solid lines in Fig. 3, where discrete points are experimental data for Tl-2212 with $B_\Phi \approx 1.5$ T.

The resulting values for the penetration depth are well behaved. This can be seen in the inset to Fig. 3 (filled triangles for this sample), where $(1/\lambda_{ab})^2$ varies linearly with temperature, in keeping with Ginzburg-Landau theory. The inset also includes results for a similar analysis of Tl-2212 with a lower proton dose, as well as values for unirradiated material that were obtained using the conventional London relation in Eq. 1. The progressive increase in $\lambda$ with the addition of CD's, which is evident in the inset, has been accounted for [14] using the theory of Wahl et al. [16]. Finally, we note that the values for the pinning energy used for the modeling in Fig. 3 are reasonable, with $U_o = (0.9 - 1.2) \times \varepsilon_o$. The relative success of the modeling and the good behavior of the associated parameters all support the essential correctness of the underlying assumptions, namely the refocusing by anisotropy and modification of the mixed state energy and equilibrium magnetization by pinning.

The pinning provided by randomly oriented columnar defects is strong, as this is a form of correlated disorder. Consequently one expects the Tl-2212 materials to conduct high levels of critical currents. To quantify this, we measured the magnetization $M(H)$ in the irreversible reversible region, also. According to the Bean critical state model, the persistent current density $J_p = 15\Delta M/r$, where $r = 2\times 10^{-4}$ cm is the average grain radius and $\Delta M = [M(H \downarrow) - M(H \uparrow)]$ (in Gauss) is the difference in magnetization measured in the decreasing ($\downarrow$) and increasing ($\uparrow$) field branches of the hysteresis loop. Results of this study are shown in Fig. 4(a) - 4(d) for materials with several defect densities, in applied fields of 0.25, 0.5, 1, and 2 T, respectively. There are several noteworthy features. First, the additional pinning increased the persistent current density by one to two orders-of-magnitude, relative to the virgin superconductor. Accompanying this was an elevation of the irreversibility line by ~ 10 – 15 K; here we define the irreversibility line $B_{irr}(T)$ by the operational criterion that the current density $J_p$ approaches the noise floor of the experiment, which is ~500 A/cm$^2$ in this case. Third, higher fields suppress $J_p$, as expected. Fourth and finally, it is qualitatively clear that the addition of defects provides diminishing returns, and a density near 1 T gives the highest current density over most of the $H,T$ phase space. The last feature is likely due to a reduction in the overall order parameter (as evidenced by a decreasing $T_c$), and the increased penetration depth that lowers the vortex line energy, as discussed below. In addition, it is possible that the presence of many nearby CD's may facilitate vortex hopping, by providing another pinning site in close proximity.

To gain further insight into the interaction of vortices with defects in these Tl-2212 materials, we studied the time dependence of the persistent current density, i.e., flux creep. With thermally activated depinning of vortices as developed by Maley [18], the decay of current density $J(t)$ with time $t$ is governed by a master rate equation [19], with

$$-dJ/dt = (J_c/\tau)e^{-U(J)/T} \qquad (4)$$

In this expression, $\tau$ is the elementary attempt time for vortex hopping, $J_c$ is the *critical* current density, and $U(J)$ is the effective pinning energy at (instantaneous) current $J$. Inverting Eq. 4 allows one to extract $U(J)$ from the measurements by varying the single unknown $\ln(J_c/\tau)$ to form a continuous curve, by piecing together the thermal decay at different temperatures. This "Maley analysis" gives the results shown in Fig. 5, which includes data for the virgin sample in the range $T$

= 5 K - 22.5 K and for the irradiated sample with $B_\Phi = 1$ T, in the range 5 K - 32.5 K. At these relatively low temperatures, the fundamental superconductive parameters vary little and the factor $\ln(J_c/\tau)$ can be taken as constant. With the logarithmic $J$-axis in Fig. 5, the linear variation (dashed line in the figure) for the virgin Tl-2212 corresponds to the dependence $U(J) = U_0 \times \ln(J_c/J)$, which arises for weak collective pinning of single vortices [19]. The experimental energy scale $U_0 = 160$ K is reasonable, as it is about ½ of the vortex pancake energy $U_{pancake} = \varepsilon_o c_0 = (\Phi_o/4\pi\lambda_{ab})^2 c_0$ where $c_0 = 1.5$ nm is the c-axis lattice parameter. For this estimate, we use the value $\lambda_{ab} = 288$ nm obtained by linearly extrapolating to $T = 0$ the data in the inset of Fig. 3. For the irradiated material, the functional dependence $U(J)$ is more complex. At the lowest temperatures (highest $J$'s), the apparent horizontal shifts and mismatch in the segments give evidence for quantum tunneling of vortices, although the effect is less pronounced than that observed in Bi-2212/Ag tapes. [8] What is clear is that adding random CD's increases the effective pinning energy at any level of $J$, relative to the defects present in the as-synthesized Tl-2212. Finally, extrapolating the $U(J)$ curves to $U = 0$ gives values for $J_c$ at $T = 0$. Combining these with the values for $\ln(J_c/\tau)$ from the "Maley analysis" provides the estimate $\tau \sim 4 \times 10^{-9}$ s for the elementary attempt time. Similar values are obtained for both materials in Fig. 5. While the accuracy is only logarithmic, this procedure does provide estimates for this difficult-to-assess quantity.

We have shown that irradiation with GeV protons leads to significant increases in vortex pinning in many high-$T_c$ superconductors. The energetic particles induce fission of Hg, Tl, Bi, ... nuclei and the recoil fragments amorphize the host superconductor, forming randomly oriented columnar defects. For the materials studied here, the transition temperature decreases linearly with proton fluence, at a rate that depends mostly on the probability per unit volume for inducing fission. The enhanced pinning increases the persistent current density and elevates the irreversibility line. For the Tl-2212 material, the addition of defects strongly affects the equilibrium magnetization. Its magnitude decreases significantly and the magnetic field dependence changes from a simple London $\ln(H)$ form (as observed for the virgin material) to a more complex "S"-shaped dependence. Using a theoretical model to account for vortex-defect interactions and invoking an anisotropy-induced refocused vortex-defect array, we modeled the field and temperature dependent magnetization using reasonable parameter values. In conclusion, irradiation with GeV protons can form randomly oriented columnar defects deep within superconducting materials, which both enhances the current-carrying performance of the material and provides interesting physical systems for investigating the interplay between vortices and correlated disorder.

We wish to thank M. Paranthaman for providing the Tl-2212 starting materials used in this study. Oak Ridge National Laboratory is managed by UT-Battelle, LLC for the U.S. Department of Energy under contract DE-AC05-00OR22725. Los Alamos National Laboratory is funded by the US Department of Energy under contract W-7405-ENG-36. Work by JGO is sponsored by FONDECYT project 1000394

**Figure captions**
Fig. 1. The dependence of $T_c$ on fluence of 0.8 GeV protons, for the several HTS materials indicated. Inset: the $T_c$ depression rate versus the product $\sigma_f (N/V)_{nuclei}$ that controls the probability of generating a random columnar defect.
Fig. 2. The equilibrium magnetization of polycrystalline Tl-2212 at 40 K, plotted against magnetic field on a logarithmic axis. Materials were irradiated with 0.8 GeV protons to create randomly oriented columnar defects with densities corresponding to the 'matching fields' shown.
Fig. 3. The equilibrium magnetization of Tl-2212 with random columnar defect density $B_\Phi \cong 1.5$ T, versus field $H$. Symbols are experimental points at the temperatures shown; lines are model calculations using Eq. 3. Inset: temperature dependence of the London penetration depth used for modeling this sample (filled triangles) and others; see text.
Fig. 4. The persistent current density versus temperature for Tl-2212 materials with defect densities $B_\Phi \cong 0, 0.08, 0.18, 1,$ and 1.5 T as indicated, in applied magnetic fields of (a) 0.25 T; (b) 0.5 T; (c) 1 T; and (d) 2 T.
Fig. 5. A 'Maley analysis' of the effective vortex pinning energy $U(J)$ versus current density $J$, for virgin and irradiated Tl-2212 measured in an applied field of 1 T.

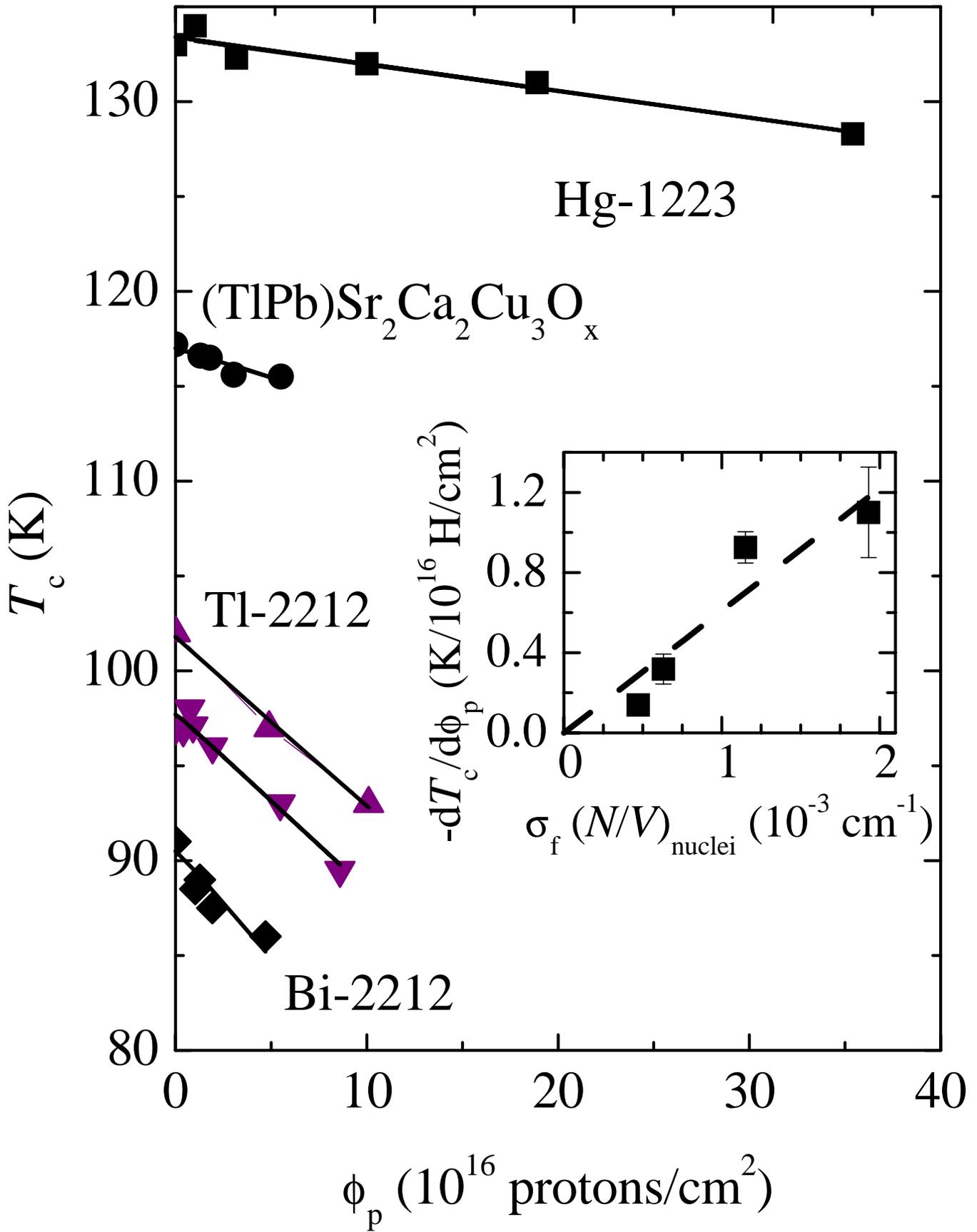

Fig. 1 VP-2 / ISS 2001 Thompson

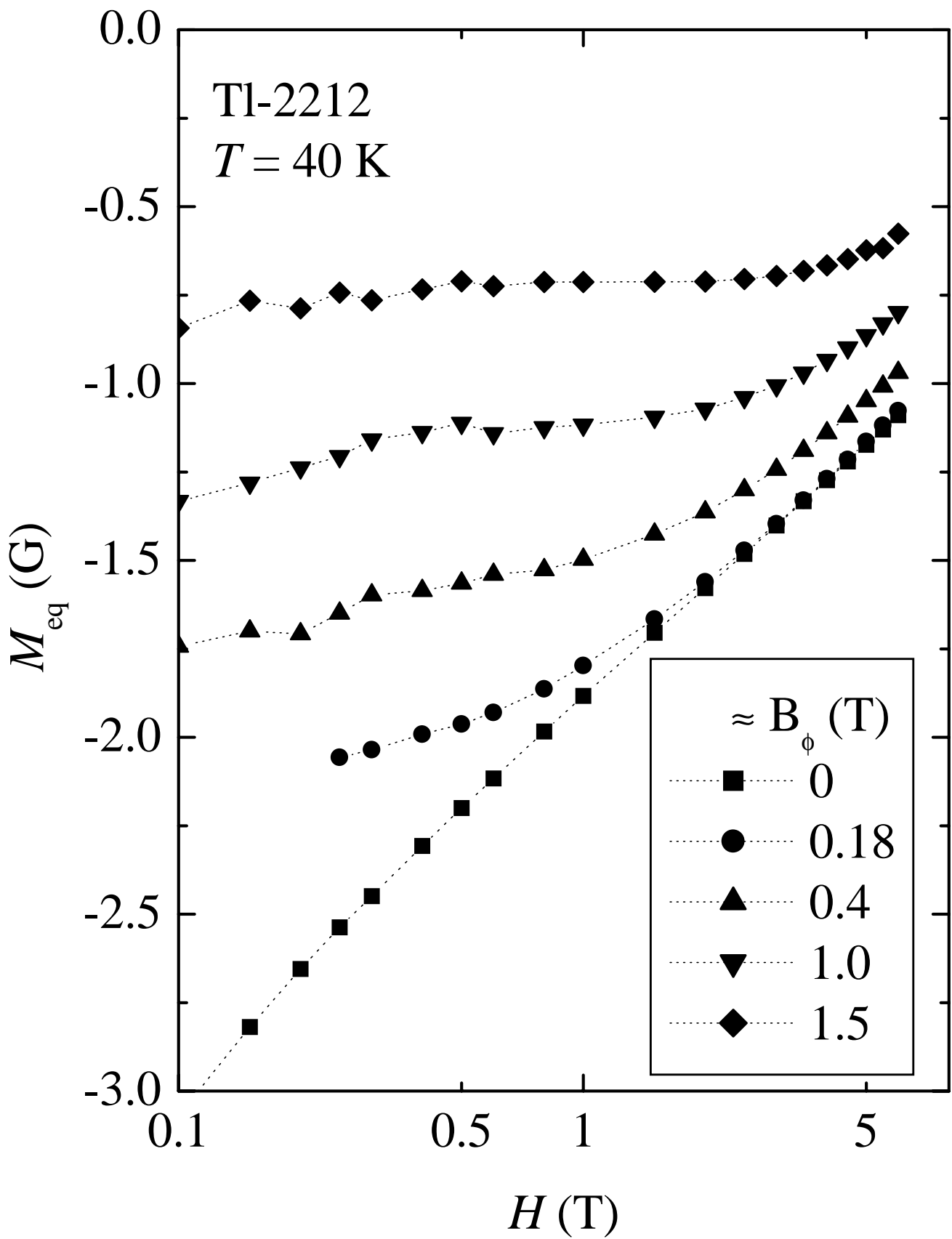

Fig 2 VP-2 / ISS 2001 Thompson

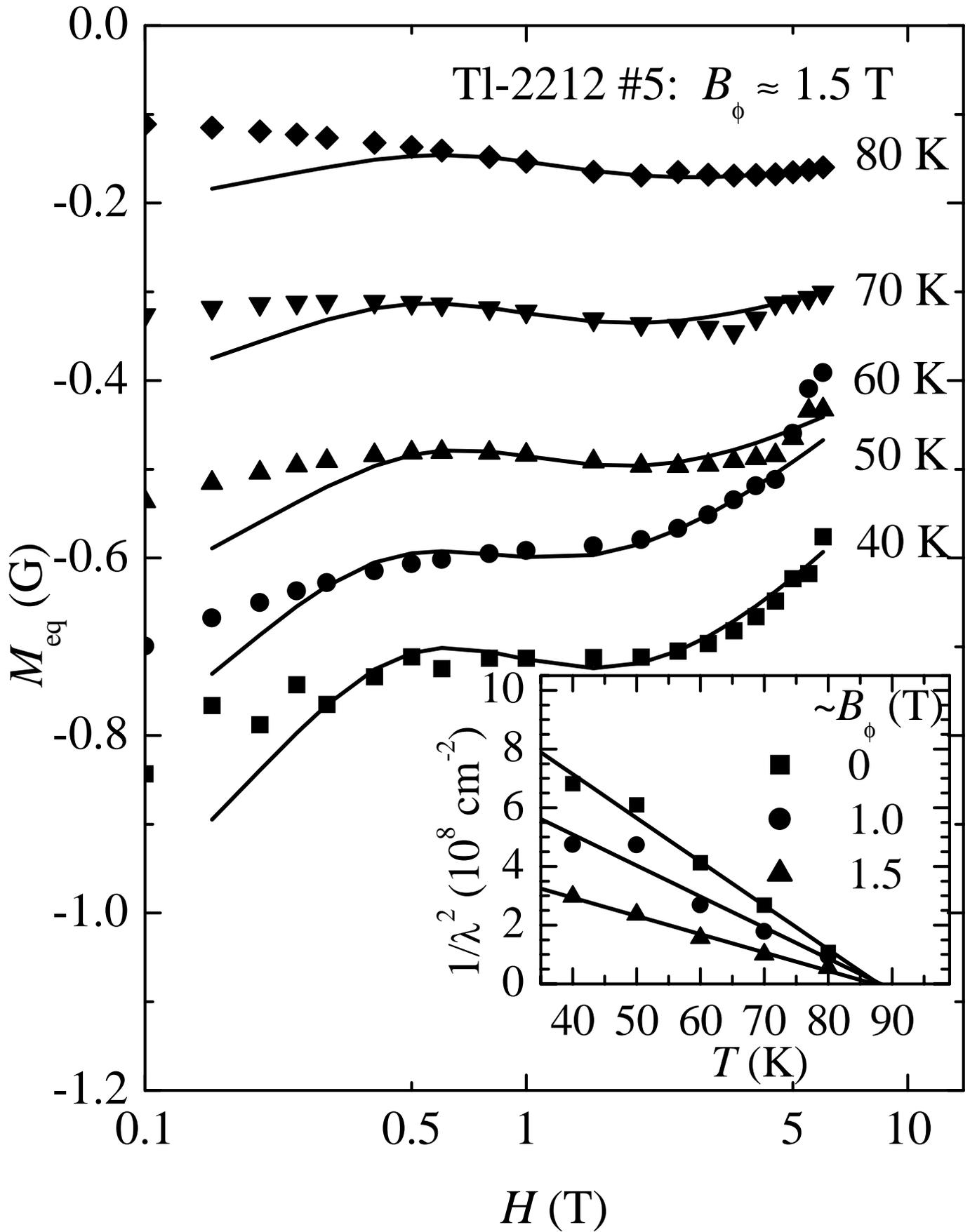

Fig 3 VP-2 / ISS 2001 Thompson

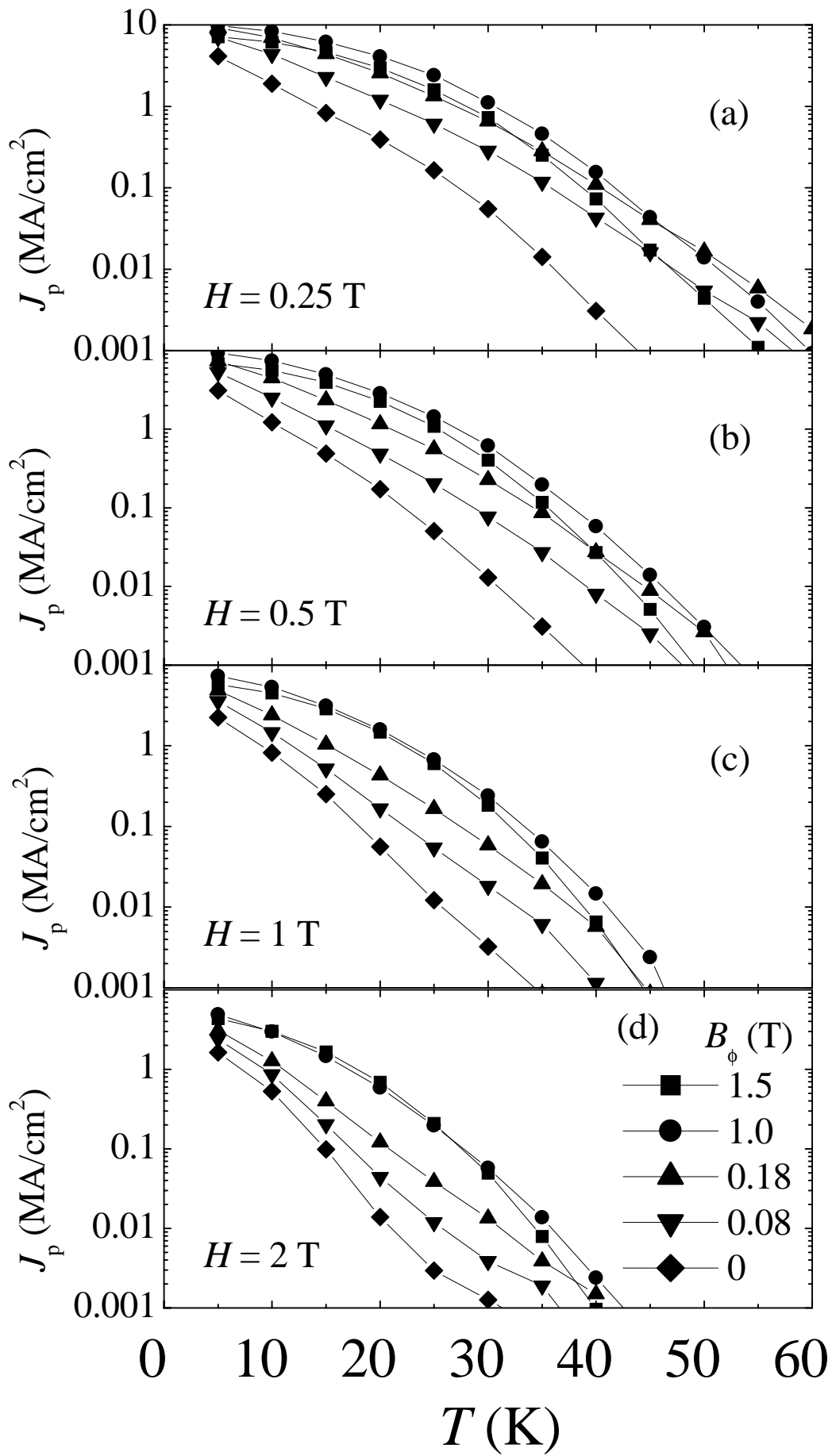

Fig 4 VP-2 / ISS 2001 Thompson

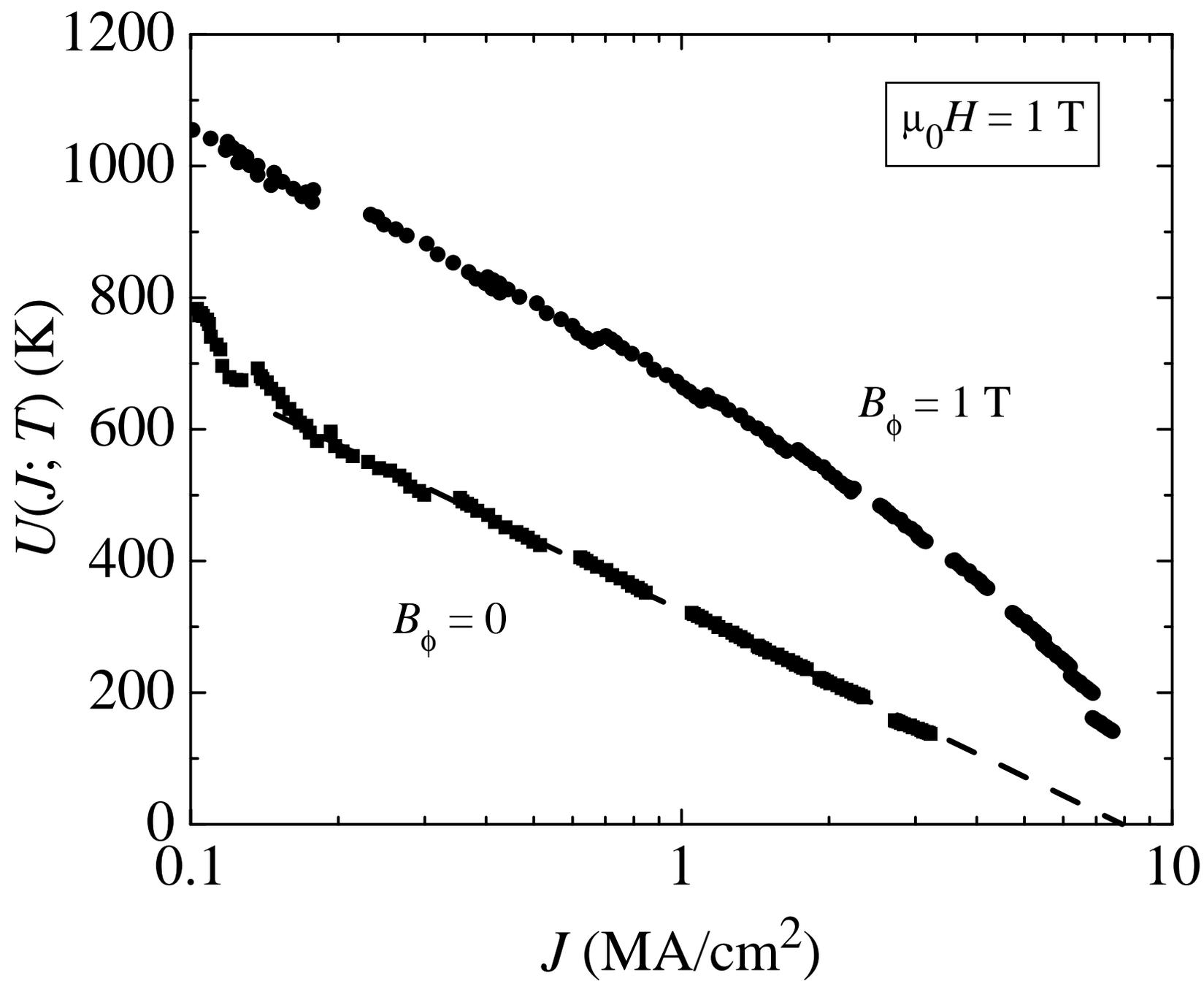

Fig 5 VP-2 / ISS 2001 Thompson